\documentclass[aps,preprint,showpacs]{revtex4}
\usepackage{graphicx}
\begin{document}
%
\title{
{\it \bf Ab initio} study of reflectance anisotropy spectra of a sub-monolayer
oxidized Si(100) surface
}
\author{Andr\'e Incze }
\affiliation{
 Istituto Nazionale per la Fisica della Materia, Dipartimento
      di Fisica dell'Universit\`a di Milano, via Celoria 16, I-20133
      Milano, Italy and 
Istituto Nazionale per la Fisica della Materia e Dipartimento di Fisica
   dell'Universit\`a \\ di Roma ``Tor Vergata'',
   Via della Ricerca Scientifica, I--00133 Roma,
   Italy }
\author{Rodolfo Del Sole }
\affiliation{
Istituto Nazionale per la Fisica della Materia e Dipartimento di Fisica
   dell'Universit\`a \\ di Roma ``Tor Vergata'',
   Via della Ricerca Scientifica, I--00133 Roma,
   Italy }
\author{Giovanni Onida }
\affiliation{ Istituto Nazionale per la Fisica della Materia, Dipartimento
      di Fisica dell'Universit\`a di Milano, via Celoria 16, I-20133
      Milano, Italy }
\date{\today}
\begin{abstract}

The effects of oxygen adsorption on the reflectance anisotropy spectrum (RAS) of reconstructed Si(100):O 
surfaces at sub-monolayer coverage (first stages of oxidation)
have been studied by an {\it ab initio} DFT-LDA scheme within a plane-wave, 
norm-conserving  pseudopotential approach. Dangling bonds and the main features of the characteristic RAS of the clean Si(100)
surface are mostly preserved after oxidation of 50\% of the surface dimers, with some visible changes: a small red shift of the first peak, and the appearance of a distinct spectral structure at about 1.5 eV. The electronic transitions involved in the latter have been analyzed through state--by--state and layer--by--layer decompositions
of the RAS. We suggest that new interplay between present theoretical results and reflectance anisotropy spectroscopy experiments
could lead to further clarification of structural and kinetic details of the 
Si(100) oxidation process in the sub-monolayer range.

\end{abstract}




%
\maketitle

\noindent 

The developments in the capability  to monitor and to control oxidation of the silicon (100) 
surface at an atomic-scale resolution
are a very important issue in semiconductor device technology.  In fact,  high quality gate insulators thinner
than 2nm are today required, due to the continuous downscaling of the metal-oxide-semiconductor (MOS) devices
(see, e.g., Ref. \cite{BaumvolSurfSci1999}).

In the last two decades the adsorption of oxygen on silicon surfaces has been extensively 
studied, from both experimental and theoretical points of view. In particular, 
the effects induced by oxidation on the optical properties of the silicon surfaces are being thoroughly 
studied 
since the mid eighties \cite{SelloniPRB33_8885_1986,KeimSurfSci1987}. It has been known
since then that O$_2$ adsorption is dissociative, and that preferred final adsorption sites 
for O atoms are Si-Si bond-bridge positions. 
More recently, it has been shown
that in the case of Si(100) the most favorable adsorption sites for O are the 
dimer-bridge, and the backbond site of the down atom of the surface dimer \cite{UchiyamaSurfSci1996,
UchiyamaPRB53_7917_1996}. The latter site is slightly favored over the former when insertion of a 
single O atom is considered. When two oxygen atoms originated by the dissociation of an O$_2$ molecule
are present, then the most favorable configuration which can be reached in a barrierless or almost
barrierless reaction (i.e. with oxygen present only on surface sites) is that with one oxygen inserted
into the down atom backbond, and the other inserted into the 
dimer bond \cite{UchiyamaSurfSci433-435_896_1999,WidjayaJCP116_5774_2002}. It is  
in fact known that, during the first stages of oxidation, O$_2$ adsorption and O migration
to the dimer backbonds proceed as barrierless (or almost barrierless) reactions, as demonstrated
by both experimental \cite{SREM1998} and theoretical works \cite{KatoPRL80_1998,KatoPRB62_15978_2000}.
Configurations with oxygen moving deeper into the second and/or third layer have been theoretically predicted \cite{KageshimaPRL_81_5936_1998}, but require the overcoming of a non-negligible energy barrier and are not considered here.  
Further Si(100) oxidation has been shown to proceed in a layer-by-layer manner, 
giving rise to an 
oscillatory effect in the surface anisotropy, which has been detected by 
"in situ" reflectance anisotropy spectroscopy measurements \cite{Yasuda2001,Matsudo-Yasuda2002}.
The phenomenon of the growth of a SiO$_2$ layer over the Si(100) surface, and the related
problem of the properties of the Si-SiO$_2$ interface,
have received most of the attention, and several calculations of structural and 
electronic properties of the interface are available \cite{PasquarelloNature1998, xxxxAPL2000}. 
However, what happens in the very first stages of oxidation (below 1 ML coverage) is still poorly known: 
oxidation of the second layer starts soon \cite{NakayimaPRB63_113314_2001}, and 
recent experimental works \cite{Matsudo-Yasuda2002, Yasuda2003} do not focus on the formation of the 
first layer, which is completed in a very short time (less than 1 second) at the experimentally used 
O$_2$ partial pressures \footnote{The experimental conditions, with an O$_{2}$ partial pressure as large as 
10$^{-2}$ Pa, lead an exposure at about 10$^{3}$ Langmuir in only a few seconds. As a consequence, the well-known surface originating negative RAS peaks at low energy disappear almost immediately (see figures 5 and 6 in reference {\protect\cite{Matsudo-Yasuda2002}}).}. 
Since reflectance anisotropy spectroscopy is a very sensitive technique, which in the case of the clean Si(100) surface 
is able to detect small structural details \cite{ShiodaPRB57_R6823_1998, RossowManteseAspnesJvac1996}, 
and --in contrast with other characterization techniques based on electrons-- can be used ''in situ'' to 
monitor the process under  manufacturing conditions, we suggest, with the present work, a possible 
application of reflectance anisotropy spectroscopy to shed light on  the structural and kinetic details of the
Si(100) oxidation process in the sub-monolayer range.

Our aim  is to study 
in a  detailed and quantitative way  the changes on the 
RAS features induced by very low concentrations of
chemisorbed oxygen on the  Si(100) surface. To this scope, a layer-resolved analysis of the most relevant 
low-energy spectral features will be presented.

\section{Theory}
The RAS is defined as the difference between the normalized 
reflectivities measured at normal incidence, for two orthogonal
polarizations of light. Throughout this
paper, the surface is perpendicular to the $x$ axis, and the $z$ direction is 
perpendicular to the Si-Si dimers of the reconstructed Si(100) surface.
RAS is hence measured, as a function of the photon energy, as $({\Delta{R_{y}}}/{R})-({\Delta{R_{z}}}/{R})$,
where R is the (isotropic) Fresnel reflectivity.
Describing the surface within a repeated slab geometry,  one can express ${\Delta{R_{i}}}/{R}$, ($i=y,z$) 
for normally incident light as \cite{manghi}:

\begin{equation}
\frac{{\Delta}R_{i}}{R}=\frac{{4\omega}}{c}Im\frac{4{\pi}{\alpha}_{ii}^{hs}(\omega)}{{\epsilon}_{b}({\omega})-1},
\label{ras_halfslab}
\end{equation}

where $i$ is the  polarization direction, $\alpha_{ii}^{hs}(\omega)$ are the diagonal terms of the 
half-slab polarizability tensor, 
and ${\epsilon}_{b}(\omega)$ is the complex bulk
dielectric function. 

If the slab has a center of inversion or a mirror plane parallel to the surface ($x{\rightarrow}-x$ symmetry, in our
case),
the imaginary
part of $\alpha_{ii}^{hs}(\omega)$ can be written, in the single--quasiparticle approximation, as:

\begin{equation}
Im[4{\pi}{\alpha}_{ii}^{hs}(\omega)]=\frac{4\pi^{2}e^{2}}{{\omega}^{2}A}\sum_{\vec{k}}\sum_{v,c}{|V_{v\vec{k}, c\vec{k}}^{i}|}^{2}
\delta(E_{c\vec{k}}-E_{v\vec{k}}-{\hbar}{\omega}),
\end{equation}

where $V_{v\vec{k}, c\vec{k}}^{i}$ are the matrix element of the velocity operator between occupied ($v$) and empty ($c$)
slab eigenstates at the point $\vec{k}$ in the surface Brillouin zone \cite{delsole1}.
"A" is the slab surface area, while $E_{c\vec{k}}$ and $E_{v\vec{k}}$ are conduction and valence energy eigenvalues, taken 
as representative of quasiparticle energies.
Neglecting the pseudopotential nonlocality \cite{noi_nonloc}, the velocity operator can be replaced by the momentum 
operator divided by 
the electronic mass, whose matrix elements $P_{v\vec{k},c\vec{k}}^{i}$ are easy to evaluate in the plane-wave basis. 
If the slab has two identical surfaces, $\alpha_{ii}^{hs}$  is simply obtained by dividing by two the polarizability
of the full slab. There are many cases, however, where one needs to single out the contributions coming from an 
individual half of the slab, or from regions situated at different depths below the surface. This problem has 
been recently discussed by some of us \cite{hogan} and by others \cite{bechstedt,monachesi}, and can be solved 
by introducing a  real-space cutoff in the definition of the matrix elements.

This method proved itself very useful for
separating surface contributions  from bulk and subsurface ones in optical spectra, and has already
been applied to the case of the clean Si(100) reconstructed surface \cite{hogan}.
In the real-space cutoff scheme, modified matrix elements $\tilde P_{v\vec{k}, c\vec{k}}^{i}$ are
needed,
which incorporate a function $\theta(x)$ switching from 1 inside the selected region to zero outside
the selected region. $\tilde P_{v\vec{k}, c\vec{k}}^{i}$ are defined as:

\begin{equation}
{\tilde{P}}^{i}_{v\vec{k},c\vec{k}}=-i{\hbar}{\int}{d^{3}\vec{r}}{\Psi}_{v\vec{k}}^{*}(\vec{r}){\theta(x)}\frac{\partial}{{\partial}r_{i}}{\Psi}_{c\vec{k}}(\vec{r}).
\label{eq:cutoff}
\end{equation}

The imaginary  part of the polarizability  is hence given by \cite{hogan}:

\begin{equation}
Im[{\alpha}_{ii}^{cut}]=\frac{2\pi{e^{2}}}{m^{2}{\omega}^{2}A}{\sum}_{\vec{k}}{\sum}_{v,c}[P_{v\vec{k},c\vec{k}}^{i}]
^{*}{\tilde{P}}_{v\vec{k},c\vec{k}}^{i}{\delta}(E_{c\vec{k}}-E_{v\vec{k}}-{\hbar}\omega),
\label{eq:alpha_cutoff}
\end{equation}

where both $\tilde{P}^{i}_{v\vec{k},c\vec{k}}$ 
and  $P_{v\vec{k},c\vec{k}}^{i}$ (the standard momentum matrix element,  
calculated without the cut-off function) appear.
Working in the reciprocal space, the evaluation of  $\tilde{P}^{i}_{v\vec{k},c\vec{k}}$ requires a double sum
over the reciprocal lattice vectors \{$\vec{G}$\}, 
 in contrast to the single sum which yields ${P}^{i}_{v\vec{k},c\vec{k}}$ \cite{hogan}, \footnote{Because of the double sum, the computation of the matrix elements using the cutoff 
technique can become time consuming, especially when a high cutoff energy is required for convergence. We have
checked convergence of the matrix elements calculation over the number of $\vec{G}$--vectors included in the sum.
We have obtained that 4000 $\vec{G}$-vectors are enough for achieving convergence, approximatively 1/5 of the initial basis size.}.
\section{computational details}
Electronic wavefunctions and eigenvalues are obtained within the local density approximation (LDA) to density-functional theory \cite{hohenberg, sham} using a plane-waves basis set. The exchange-correlation energy is evaluated according to 
the Ceperley and Alder
results \cite{alder} as parametrized by Perdew and Zunger \cite{perdew}. 
The ion-electron interaction is represented by 
norm-conserving pseudopotentials. 
Special attention was devoted to the generation of a good norm-conserving pseudopotential for oxygen, in order to
achieve an high transferability without requiring too many plane waves for convergence. To this aim, we adopted
the Hamann scheme \cite{hamann}, and optimized the core radii in order to find the best compromise between the 
basis set convergence and transferability. The latter was checked not only against logarithmic derivatives,
but also performing explicit atomic calculations in several excited configurations.
More extensive pseudopotential tests have been performed on three small molecules (SiO, H$_{2}$SiO, Si$_{2}$O) 
and on the $\alpha$-quartz crystalline phase of silica, in order to check the convergence of their structural 
and electronic properties with respect to the basis set (number of plane waves). As a result, the theoretical length
of the Si-O bond is found to converge (within 1\%) already at a 30 Ry cutoff for all three molecules. The
theoretical values of 1.51 {\AA} (SiO) and 1.52 {\AA} (H$_{2}$SiO) compare well with the experimental ones
(1.51 {\AA} \cite{vecchia_referenza_17} and 1.515 {\AA} \cite{silanone1} respectively). Concerning $\alpha$-quartz,
the calculated lattice constant at 30 Ry is 4.87 {\AA}, to be compared with the experimental value of 4.916 {\AA} 
\cite{expt_alpha_quartz}.

The Si(100) surface was simulated by a repeated slab of twelve silicon layers 
and four layers of vacuum. The surface unit cell has been choosen as a (2x2) one, for
computational convenience, in order to be able to consider both the (2x1) and p(2x2)
reconstructions within the same cell. 
First, the clean surface structure has been determined by a full structural optimization
 (keeping only the central four Si layers as fixed), using  
the Broyden-Fletcher-Goldfarb-Shanno minimization algorithm as implemented in
the ABINIT code \cite{abinit} until the residual forces acting on each atom are less than 0.01 eV/{\AA}.
The resulting structure was p(2x2) reconstructed in agreement with previous results \cite{RamstadPRB_51_14504_1995},
with a dimer buckling of 0.79 {\AA}. The ground state of the c(4x2) reconstruction, requiring a different surface unit cell, has also been considered for comparison.

From the structural point of view, the difference between c(4x2) and p(2x2)
is very tiny: in both cases one has rows of buckled Si dimers, and the
buckling (at difference with the (2x1) reconstruction) {\it alternates} along
those rows. The only difference between c(4x2) and p(2x2) consists in the
buckling alternance {\it in the direction perpendicular to the rows}.

Considering that, as shown in Ref. \cite{palummosi100}, the dimer-dimer interaction
is much larger along the rows than between adjacent rows
very small energetic and spectral differences between c(4x2) and p(2x2) are to be
expected.
In fact, from the theoretical point of view, the stabilities of the p(2x2) and c(4x2)
reconstructions are essentially identical \cite{RamstadPRB_51_14504_1995}.
Recently, {\it ab initio} calculations have shown that it is possible to induce
the formation of p(2x2) domains by electric fields or charge injection
\cite{SeinoPRL_93_036101_2004}.

Experimentally, the coexistence of p(2x2) and c(4x2) domains at very low temperature
is supported by recent low-temperature non contact atomic force microscopy data, which
show that almost 12 \% of the ordered Si(100) surface is p(2x2) reconstructed
\cite{UozumiSurfSci_188_279_2002}. Hence, the p(2x2) cell can be taken as a realistic
 model of the clean Si(100) surface.

Two O atoms were then adsorbed on each of the two slab surfaces (in order to preserve the inversion 
symmetry), which corresponds to a 0.5 ML oxidation.
Based on previous {\it ab initio} calculations of oxygen adsorption on the 
Si(100) surface \cite{UchiyamaSurfSci433-435_896_1999,WidjayaJCP116_5774_2002}, we 
have chosen the most stable configuration that can be reached in a 
barrierless or almost barrierless dissociation of the molecule: one O atom is in 
bridge position on a surface Si-Si dimer, and the second O atom also in bridge position
on a nearest Si-Si backbond. Many previous theoretical results \cite{UchiyamaSurfSci1996, 
UchiyamaPRB53_7917_1996, KatoPRL80_1998} have shown that: i) the preferred adsorption site of
an isolated O atom is a Si-Si backbond; ii) the backbond of the "down atom" of the dimer is strongly
 preferred over the "up atom" backbond (at the point that if adsorption occurs on the backbond of the "up atom",
this is sufficient to induce locally the reversing of the buckling); iii) when an 
O$_2$ molecule adsorbs, it dissociates leaving \cite{KatoPRB62_15978_2000} one oxygen atom into the
Si-Si surface dimer bond, and the remaining atom into the "down atom" backbond. We consider hence the configuration
denoted by (h) in Ref. \cite{WidjayaJCP116_5774_2002}.
The same structure is also considered (denoted by "A") in Fig. 1 (a) of Ref. \cite{KageshimaPRL_81_5936_1998}.
This structure has been selected among the several other
total energy minima for two oxygen atoms on Si(100), because it can be immediately reached after
O$_2$ dissociation (at difference with structures
"D" and "E" of Ref. \cite{KageshimaPRL_81_5936_1998}) and can efficiently model the effects of "breaking"
one of the surface Si-Si dimers.

\section{Results}

\begin{figure}[htbp]
\centering
\includegraphics[scale=0.75]{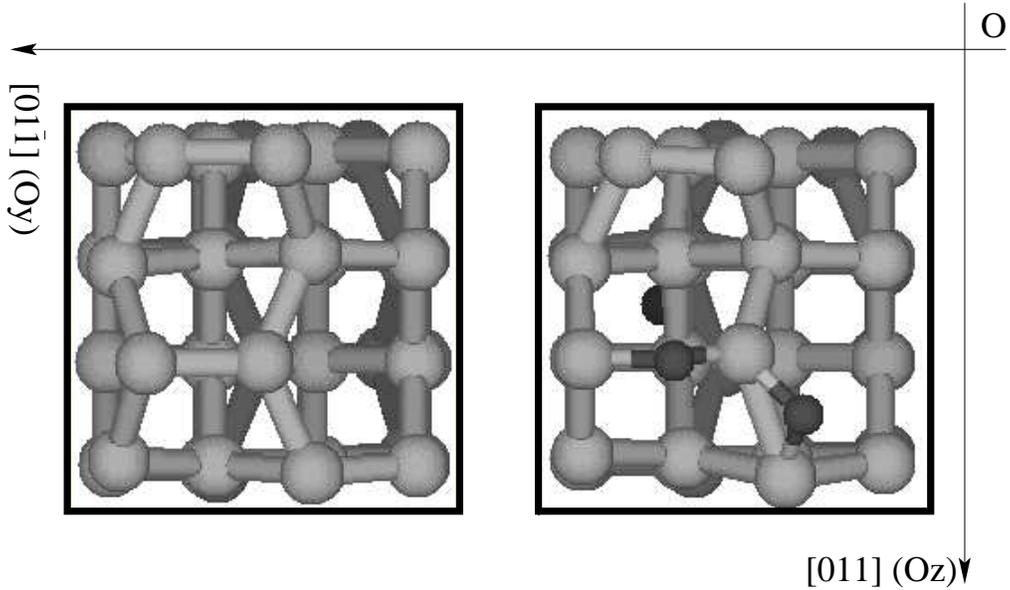}
\caption{Top view of the slab after structural optimization, for the clean (left)  and oxidized (right) Si(100)--p(2x2) 
surfaces. The $x$ axis is perpendicular to the surface, and Si-Si surface dimers are oriented along 
the $y$ axis.}
\label{fig:topfullrelaxed}
\end{figure}

After oxygen adsorption on the dimer and backbond bridge sites (Fig. \ref{fig:topfullrelaxed}), and a new full structural
relaxation, one of the surface Si-Si dimers is broken  (see Tab. \ref{tab:structural}: the 
dimer length after oxidation becomes 3.06 {\AA}). Concerning the length of the oxidized 
Si backbond, we obtain 
2.53 {\AA} after O adsorption, to be compared with the 2.60 {\AA} reported in Ref. \cite{UchiyamaSurfSci1996}.
The structural relaxation remains limited to the immediate neighbouring of the O atom, since
subsurface layers (2nd and 3rd silicon layer) are not affected appreciably.
The buckling of the  non-oxidized dimer, which is expected to be stabilized by O adsorption \cite{UchiyamaPRB53_7917_1996}, increases only very slightly, passing from  0.79  to  0.81 {\AA}.

To compute optical spectra, the ground state calculation must be followed by a calculation of bandstructure
energies and wavefunctions over a dense mesh in the irreducible wedge of the Brillouin zone, for
both occupied  and empty states. To this aim, we calculated Kohn--Sham eigenvectors and eigenvalues 
using the Arnoldi algorithm \cite{arnoldi}, for all states up to 12 eV above the highest occupied state
(i.e., about 250 empty states above the 108 filled ones) in each $\vec{k}$--point.

\begin{figure}[htbp]
\centering
\begin{tabular}{cc}
\resizebox{83mm}{!}{\includegraphics{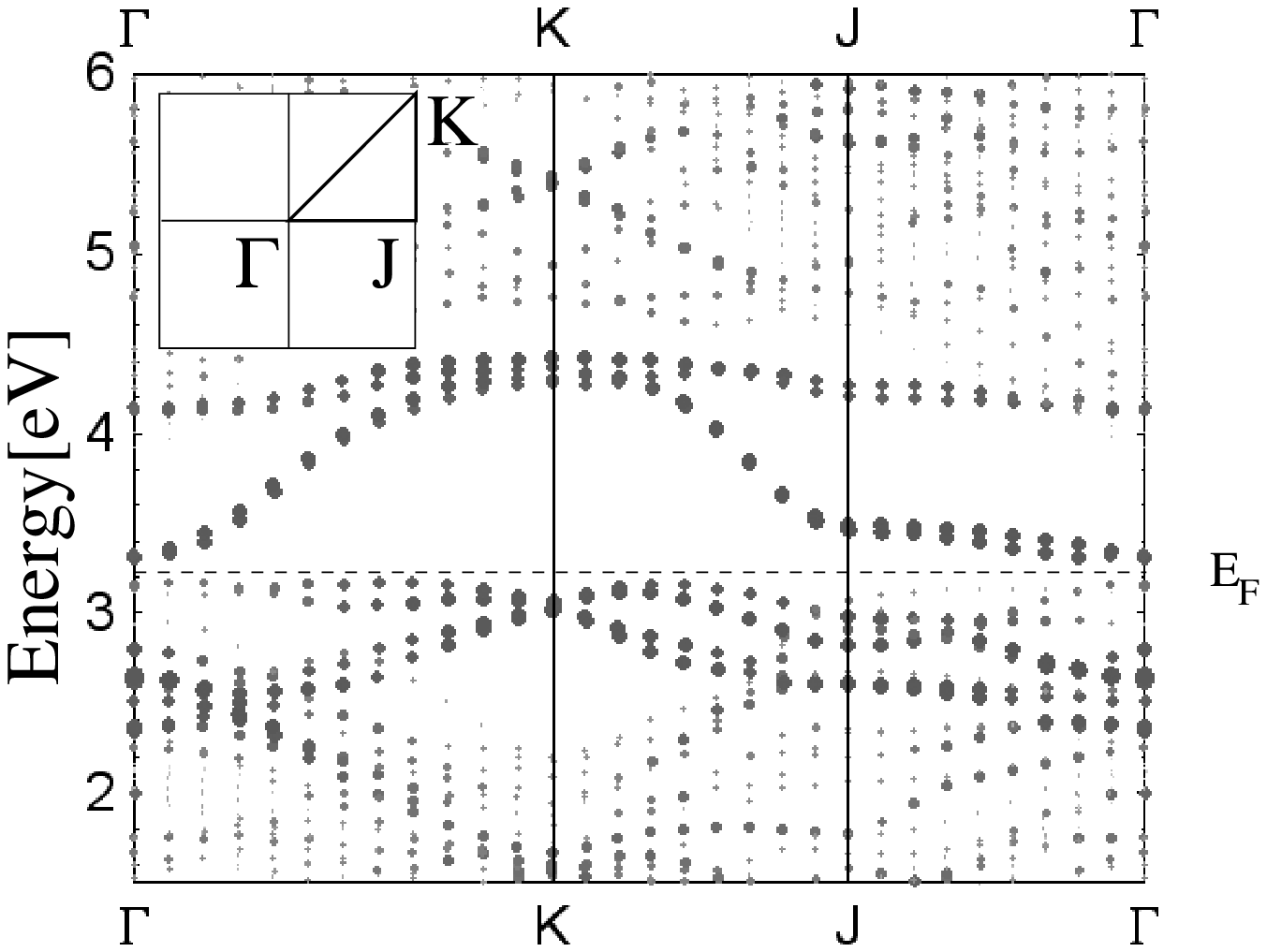}} &
\resizebox{80mm}{!}{\includegraphics{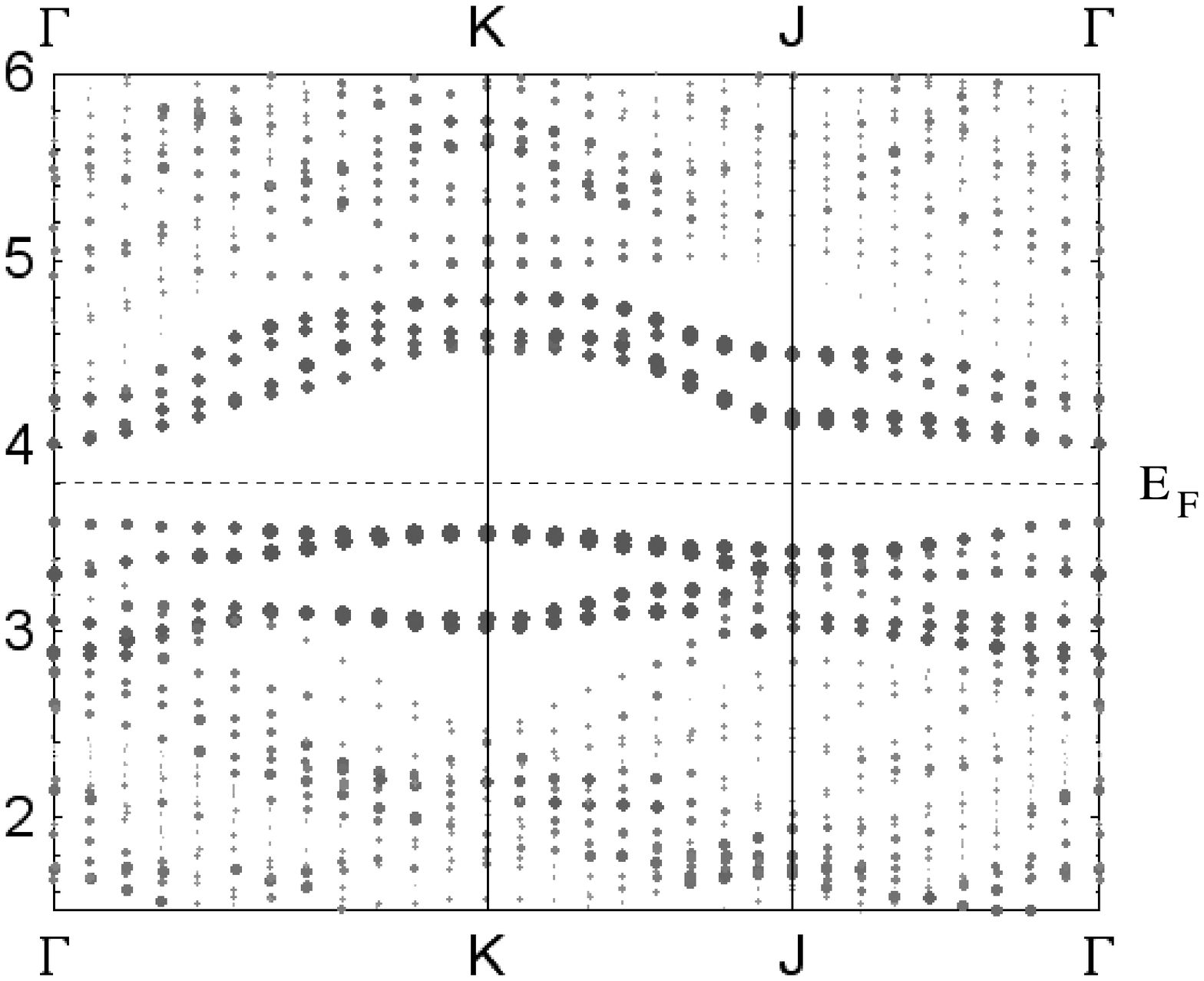}} \\
wavevector & wavevector \\
\end{tabular}
\caption{Surface bandstructure for Si(100)-$p(2\times2)$, clean (left) and 0.5 ML oxidized (right). Bold dots
are used for states which are spatially localized at the surface (integral of $|{\Psi}_{n\vec{k}}(\vec{r})|^2$ over the 2 topmost layers larger than 0.5). The Fermi level has been taken in the middle of the surface band gap.}
\label{fig:bandstructures}
\end{figure}

Fig. \ref{fig:bandstructures} shows the computed bandstructure, near the Fermi level, for both the clean and the 
oxidized surfaces, along the $\Gamma{\rm KJ}\Gamma$ path in the irreducible wedge
of the surface Brillouin zone. For the clean surface, results are in excellent agreement with
previous {\it ab initio} DFT-LDA calculations \cite{RamstadPRB_51_14504_1995} on the Si(100)--p(2x2) surface. 
We note the strong dispersion along $\Gamma{\rm K}$ and ${\rm KJ}$ for the surface bands
arising from the dangling bonds which 
form $\pi$ (bonding) and $\pi^{*}$ (antibonding) states. This large dispersion is
associated with a non-negligible interaction between adjacent silicon 
dimers in the direction perpendicular to the dimer axis \cite{palummosi100, kress} (this is
the direction along which the p(2x2) and c(4x2) surfaces are identical).

After oxidation, the direct gap at $\Gamma$ increases from 0.2 to 0.4 eV. However, surface
states are still localized on DBs, as can be seen from a plot (not shown) of their charge densities.
Even the oxidized dimers show distinct filled and empty DB-like surface states. 
The band dispersion in the direction perpendicular to dimers becomes smaller, which can be
explained by a weakening of the 
interaction between adjacent dimers (in our case, an oxidized and
a non oxidized one), as a consequence of oxidation. Oxidation also lifts the 
initial degeneracy of surface bands at
the ${\rm K}$ corner of the surface Brillouin zone.

\begin{figure}[htbp]
\centering
\begin{tabular}{cc}
\resizebox{90mm}{!}{\includegraphics[angle=-90]{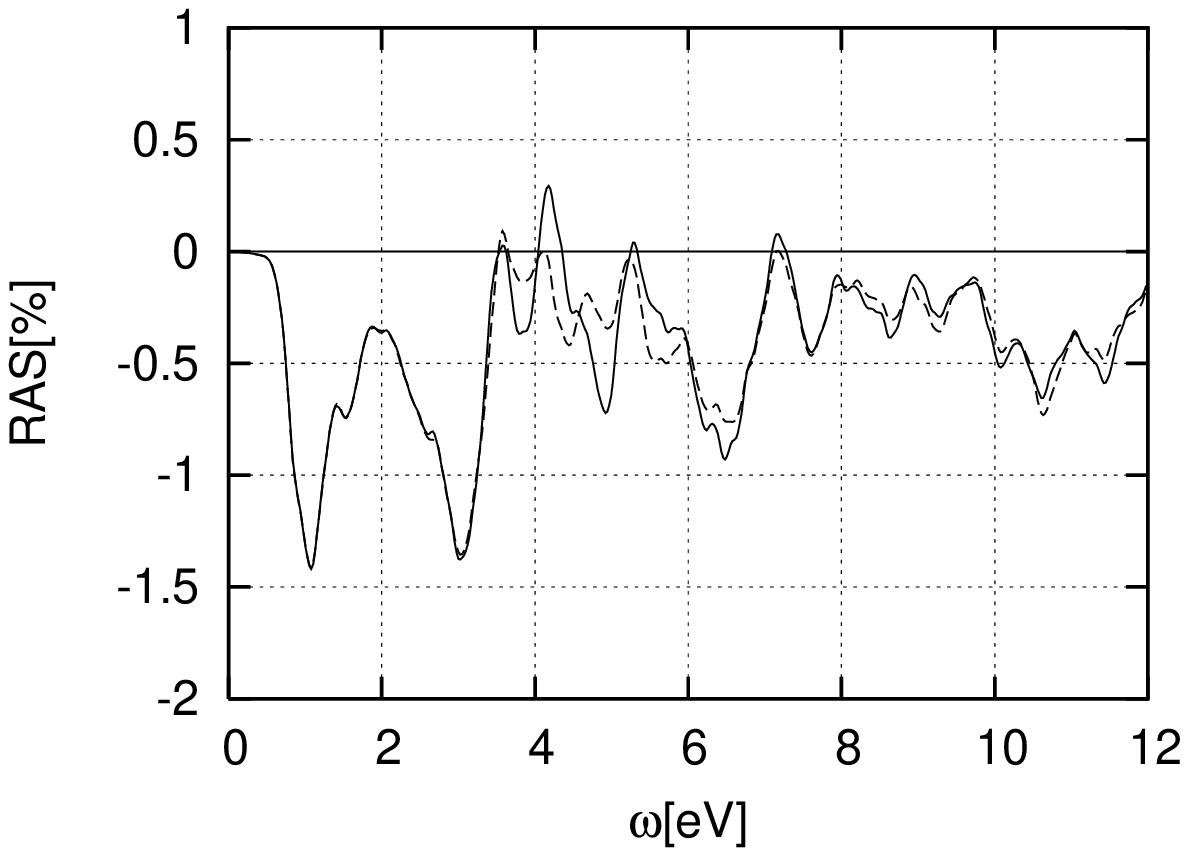}} &
\resizebox{90mm}{!}{\includegraphics[angle=-90]{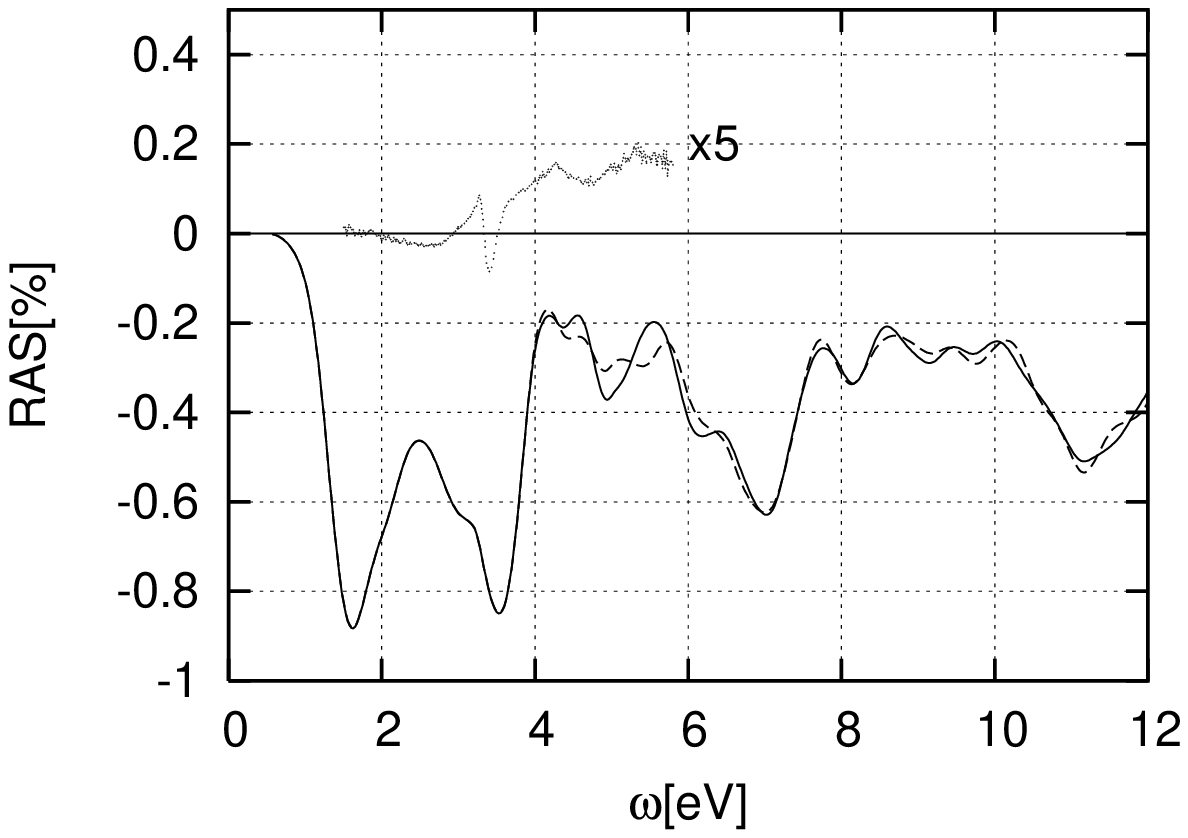}} \\
\end{tabular}
\caption{RAS of the 0.5 ML oxidized Si(100)--p(2x2), and its convergence with respect to the
Brillouin zone sampling. Left: spectra calculated with the 9x9 and 18x9 grids (continous and
dashed lines respectively), using a gaussian broadening of 75 meV. Right: spectra calculated with
the 11x11 and 18x9 grids (continuous and dashed lines respectively), using a gaussian broadening
of 250 meV. In the same panel, the lowest-exposure available experimental
data from reference \cite{Yasuda2001} are reported for comparison: however the experimental
exposure level ($\approx$ 60 s at 10$^{-2}$ Pa) is nominally much higher than that yielding
to a sub-monolayer coverage. Notice that, to facilitate comparison, in this panel the theoretical
spectra have been blue shifted by 0.5 eV to correct for the LDA error (scissor operator).}
\label{fig:convergence}
\end{figure}

Optical spectra have been computed according to Eq. (\ref{eq:alpha_cutoff}), 
using increasingly large sets of $\vec{k}$--points (up to 162). The following Monkhorst--Pack
 \cite{monkhorst},  
$\vec{k}$--point meshes were considered in the upper half of the (2x2) Surface Brillouin Zone: 5x5, 7x7, 9x9, 11x11 and 18x9.
In the energy window 0-3.5 eV, the 9x9 mesh is sufficient to achieve convergence, superimposing  a small Gaussian
broadening (75 meV) to the calculated spectrum. Similarly, in the range between 6 and 12 eV, the
11x11 mesh is sufficient, with the same small broadening.
The intermediate energy window, between 3.5 and 6 eV, is the most slowly convergent one. Indeed, a fully
converged RAS in this region can only be obtained by increasing the broadening up to 250 meV,
with the largest $\vec{k}$--point sets used (11x11 and 18x9). These results are summarized in Fig. \ref{fig:convergence}. In
the right panel, the lowest-exposure experimental data from reference \cite{Yasuda2001} are also reported for comparison. 

\begin{figure}[htbp]
\centering
\includegraphics[angle=-90,scale=0.7]{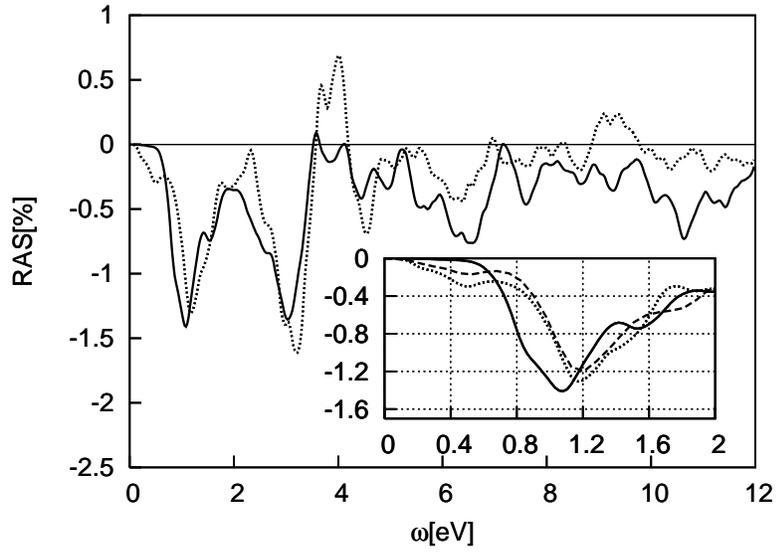}
\caption{Effects of the 0.5 ML oxidation on the RAS of Si(100)--p(2x2): clean surface (dotted line) versus
oxidized one (full line). In the inset, the low-energy region is expanded. Besides the oxidized surface (full line)
and the clean Si(100)--p(2x2) (dotted line), we also compare the RAS of the clean Si(100)--c(4x2) reconstruction (dashed 
line). A gaussian broadening of 75 meV has been used.}
\label{fig:ras.clean.oxidized}
\end{figure}

Comparing the RAS of the oxidized and clean surfaces up to 1.8 eV two differences come into evidence: 
i) after oxygen adsorption, 
there is a redshift by 0.2 eV of the main transition peak of the clean surface, and ii) a new structure appears between 1.4 and 1.75 eV (see Fig. \ref{fig:ras.clean.oxidized}). However, the overall RAS lineshape does not undergo qualitative changes.
In the inset we also plot the RAS of the clean Si(100)--c(4x2) reconstruction, which below 1.6 eV is practically identical to that of p(2x2) \footnote{The slight differences between the present results and those of Ref. {\protect\cite{palummosi100}} are due to a present better convergence in the $\vec{k}$-points sampling.}. 

\begin{figure}[htbp]
\centering
\includegraphics[scale=0.5]{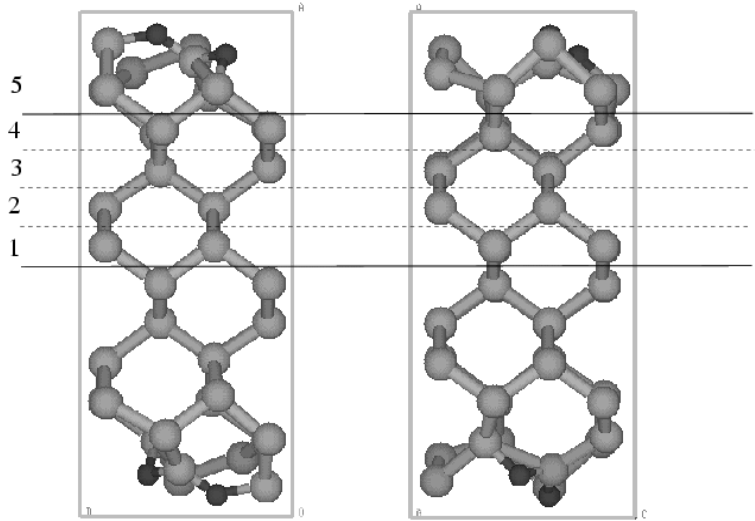}
\caption{Side view of the slab, and the slicing used for the layer--by--layer spectral decomposition performed
according to the method of Ref. {\protect\cite{hogan}}. Left: view axis perpendicular to surface dimers. Right:
view axis parallel to surface dimers. Darker spheres are used for oxygen, grey ones represent silicon atoms. 
The oxidized region  is fully enclosed in slice 5.}
\label{fig:slices}
\end{figure}

\begin{figure}[htbp]
\centering
\includegraphics[angle=-90,scale=1.0]{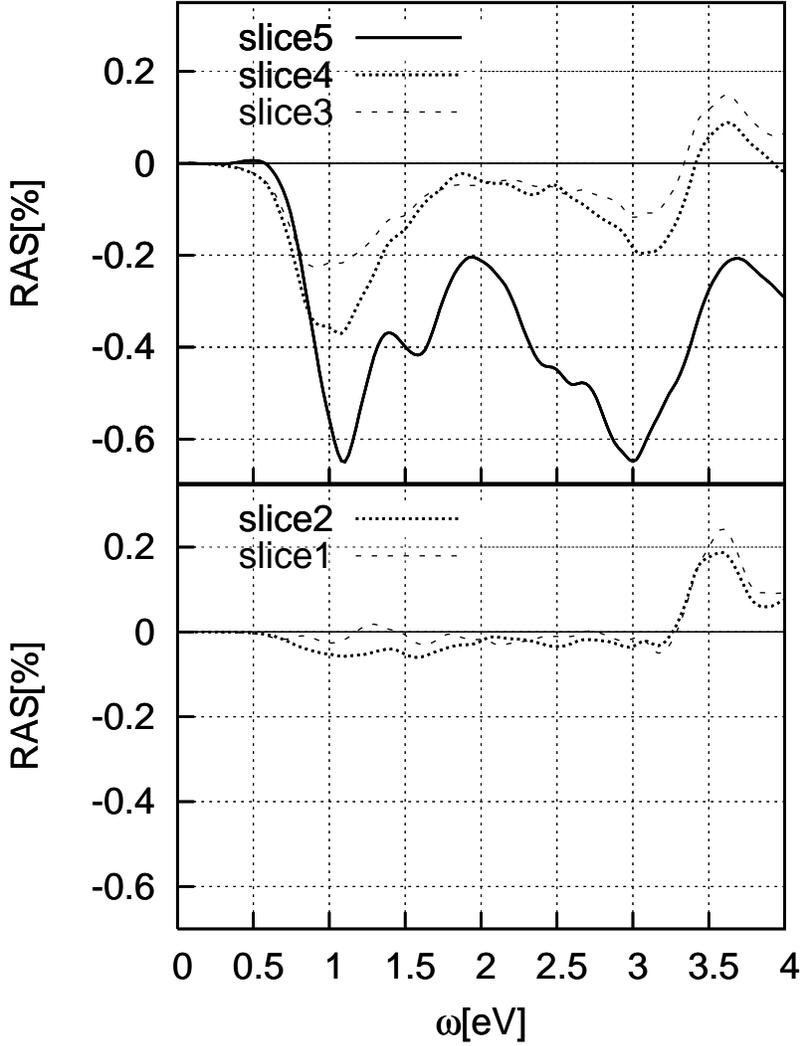}
\caption{Results of the layer-by-layer spectral decomposition, according to the slicing displayed in Fig.
{\protect\ref{fig:slices}}. The spectral feature at about 1.5 eV, recognized as an oxygen signature (see text),
comes from the topmost slice only. In contrast, the main peaks at about 1 eV and 3 eV also include contributions
from subsurface states (slice 3 and 4).}
\label{fig:cutoff_spectra}
\end{figure}

In order to clarify the origin and nature of oxygen-induced 
modifications, we have performed a layer-by-layer analysis of the RAS according to Eq. (\ref{eq:cutoff})
and Eq. (\ref{eq:alpha_cutoff}), applying the real-space cutoff method described in Ref. \cite{hogan}. 
This allows one to quantify the contribution to the RAS originated in the different surface and subsurface regions. 
We divided our slab in five slices: the first four (starting from the middle of the slab) containing only silicon atoms, and the last one including surface
dimers, oxygen atoms, and the subsurface Si atoms bonded to the surface dimers, as shown in Fig. \ref{fig:slices}. 
For the last slice, the cutoff region extends up to a distance of about 1 {\AA} above the surface.
The resulting spectra are displayed in Fig. \ref{fig:cutoff_spectra}: as expected, the reflectance anisotropy signal coming from 
slice 1 and slice 2 is almost zero up to 3.0 eV (i.e., in the whole region below the bulk Si direct gap),  
coherently with the fact that they are bulk representative. The presence of the surface starts to be felt 
in the 3rd slice, where the two main negative RAS peaks, characteristic of the clean surface,
start to appear (at about 1 and 3 eV, respectively).
Slice 4 gives a contribution very similar to slice 3, with a larger strength of the 1.0 eV
peak. 
The latter peak is also strongly present in the 
topmost slice contribution, together with the 3 eV one. 
An analysis of the localization of electronic states involved in
strongest dipole allowed transitions for the 1.0 eV peak show almost no contributions 
from electronic states involving O atoms.
Indeed, the states which originate this peak are valence states corresponding mainly to
dangling bonds localized on the upper Si atom of the oxidized dimer, and conduction
states corresponding mainly to $p_x$-like orbitals of the lower silicon atom 
of the non-oxidized dimer.
However, besides the two main peaks at 1 and 3 eV, 
the topmost slice also contributes with a new structure, located at about 1.5
eV, which does not appear in  slice 3 and slice 4 contributions.
Empty states involved in the transitions
responsible for this feature are in fact found to be strongly localized on 
the lower Si atom of the Si-O-Si bridge (broken dimer), and on the O atoms themselves ($p$-like orbitals),
so this feature represents a sort of oxygen signature.
 
\begin{figure}[htbp]
\centering
\includegraphics[scale=0.7]{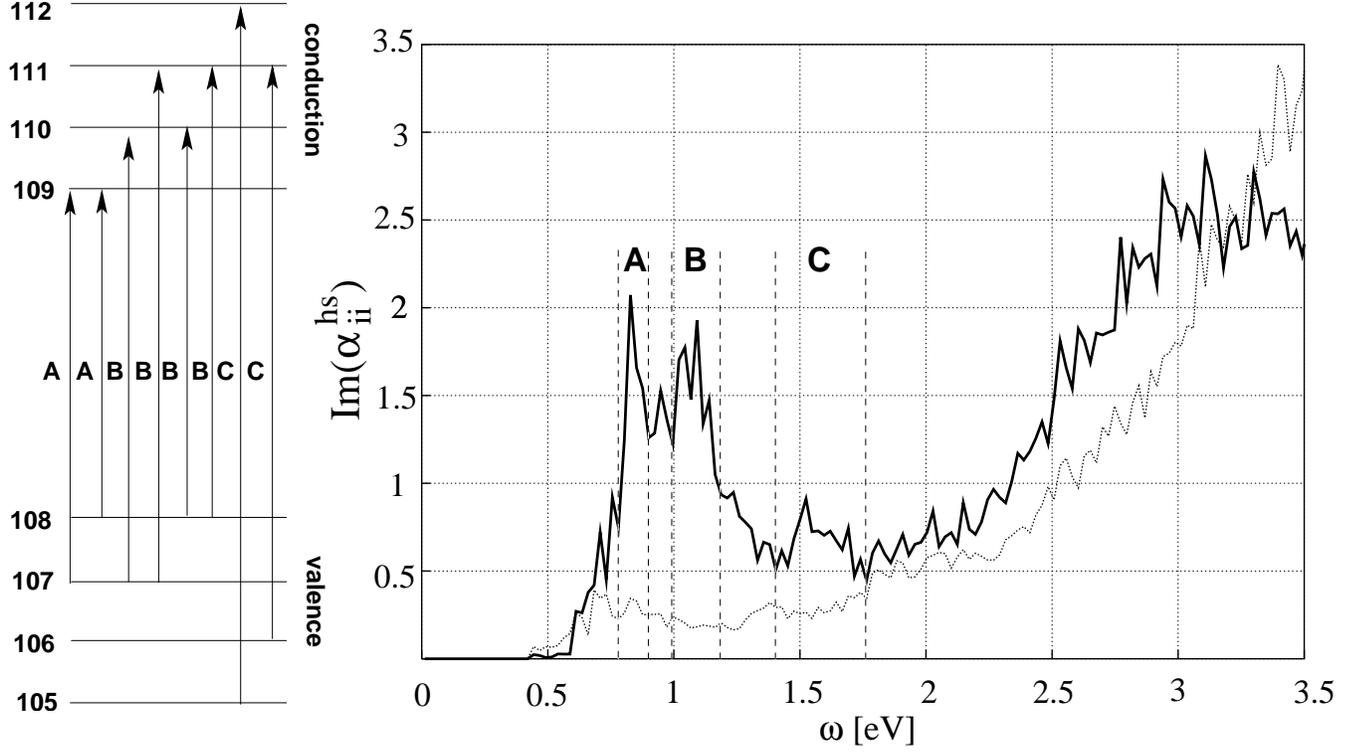}
\caption{Contributions of selected slab states involved in the strongest optical transitions for Si(100)--p(2x2):O, below 1.7 eV. Three energy windows, corresponding to the main spectral features in the imaginary part of the half--slab polarizability
tensor $Im[\alpha_{ii}^{hs}]$, are studied: A ([0.8,0.9] eV), B ([1.0,1.15] eV), and C ([1.40,1.75] eV). The full and
dotted lines correspond to light polarization along the $z$ and $y$ directions, respectively. On the left, we represent
schematically the slab states, below and above the Fermi level, and their originated transitions which carry the 
largest oscillator strength, and anisotropy, in regions A, B and C.}
\label{fig:select.trans1}
\end{figure}

In order to obtain deeper insight on the electronic states involved in the main spectral structures,
we have singled out, from the whole summation appearing in Eq. (\ref{eq:alpha_cutoff}), the contributions 
which give the largest oscillator strength and the larger anisotropy, restricted to well-defined energy
windows centered on the main low-energy peaks of the imaginary part of the slab polarizability tensor. 
To this aim, we have selected three energy windows, labeled with A, B and C (Fig. \ref{fig:select.trans1}).
based on the plot of the imaginary part of the half-slab polarizability
for light polarization perpendicular ($z$) and parallel ($y$) to the silicon dimers. 
In each region, the contribution coming from each $\vec{k}$--point and band pairs ($vc$) have been sorted
on the basis of two criteria: the value of the squared modulus $|P^z_{v\vec{k},c\vec{k}}|^2$, and the value
of the anisotropy calculated as $({|P^z_{v\vec{k},c\vec{k}}|^2 - |P^y_{v\vec{k},c\vec{k}}|^2})/({|P^z_{v\vec{k},c\vec{k}}|^2+|P^y_{v\vec{k},c\vec{k}}|^2})$. As a result, only the uppermost 4 valence states (labeled: 105--108) and the lowest 4 conduction states (labeled: 109--112) are found to be the main responsibles for the strongest optical transitions, with 107 $\rightarrow$ 109 and 108 $\rightarrow$ 109 contributing to peak A ([0.8,0.9] eV), and 107+108 $\rightarrow$ 110+111 contributing to
peak B ([1.0,1.15] eV), while  peak C ([1.40,1.75] eV) is due to the transitions 106 $\rightarrow$ 111 and 105
$\rightarrow$ 112, as illustrated in Fig. \ref{fig:select.trans1}.

\begin{figure}[htbp]
\centering
\includegraphics[angle=-90,scale=1.0]{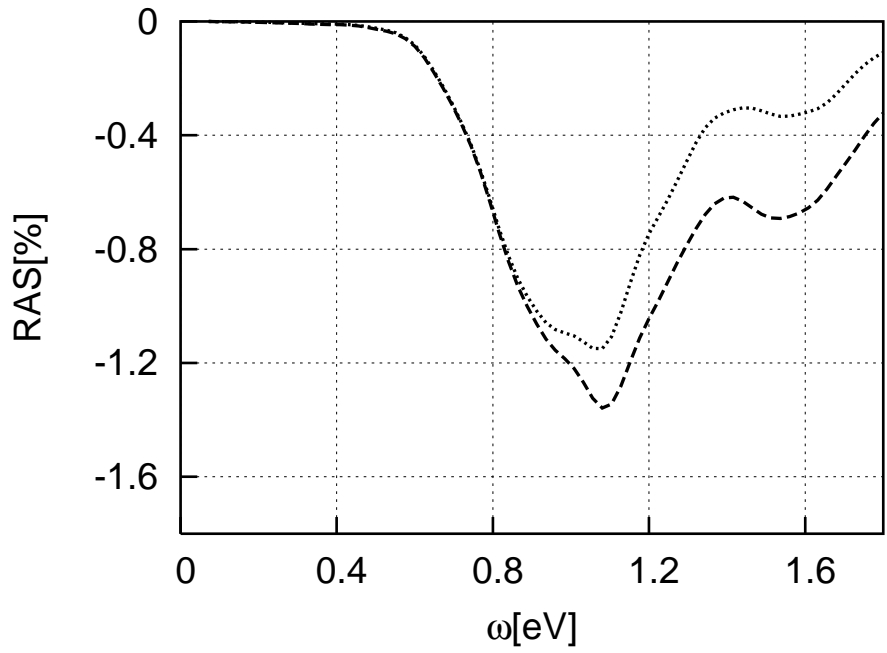}
\caption{Contribution of the eight surface states singled out in Fig. {\protect \ref{fig:select.trans1}} 
to the Si(100)--p(2x2):O RAS. Dotted line: only 4+4 bands included in the summations of 
Eq. ({\protect\ref{eq:alpha_cutoff}}); dashed line:  full calculation with 108+250 bands.}
\label{fig:select.trans2}
\end{figure}

This result is confirmed by the comparison of the full RAS with a spectrum computed including 
only this set of 4 valence and 4 conduction states (i.e., based on just 16 out of 27216 $v$-$c$ transitions): 
in the energy window between 0 and 1.8 eV all main features
of the RAS are reproduced, apart for small difference in intensities (Fig. \ref{fig:select.trans2}).

\section{CONCLUSIONS}
In conclusion, we have studied the very first stage of oxidation of Si(100)--p(2x2), at a coverage of 0.5 ML,
showing  that oxidation of one of the two Si--Si dimers of  the p(2x2) unit cell does not change
dramatically the RAS shape, despite the breaking of the dimer and the large flattening of the lowest 
conduction band, which reflects the
weakening of the dimer--dimer interaction along the direction perpendicular to the dimer axis. In particular,
the negative peak at 3.7 eV does not disappear (see footnote 42).
However, oxidation of one dimer gives rise to distinguishable effects on the RAS in the lowest energy
region (0 to 1.8 eV), which can be understood in terms of transitions between the four highest valence bands
and the four lowest conduction bands of the slab, localized essentially on dangling bonds of silicon
atoms, belonging to both types of dimers (oxidized and non-oxidized).
These effects are essentially summarized by a redshift of about 0.2 eV of the first RAS negative peak,
disappearance of the small structures below 1.0 eV, typical of the clean surface, and appearance of 
a new structure at 1.5 eV, which can be seen as an "oxygen signature". The latter can be  clearly separated
from extra artificial oscillations induced by the discreteness of the Brillouin zone sampling, 
when a sufficiently  large number of $\vec{k}$--points is used.
The shift of the first  negative peak is strongly linked to the structural 
surface relaxation which follows oxygen adsorption, and becomes  larger if the relaxation is
not complete. A layer-by-layer analysis of the RAS shows that the 1.5 eV structure is extremely 
surface--localized (topmost 2.64 {\AA}), while the origin of the two main negative peaks also extends 
somehow to the subsurface region (topmost 5.42 {\AA}). 
We predict that the appearance of this feature should be detectable in reflectance anisotropy spectroscopy experiments 
on Si(100) in the sub-monolayer range. Our theoretical predictions could be experimentally assessed by reducing the
oxygen exposure to a value of the order of 1-10 Langmuir, to be compared with the $\sim$ 10$^{3}$ Langmuir of presently
available exprimental data.

\section{ACKNOWLEDGEMENTS}
We acknowledge the European Community for financial support under the
NANOQUANTA project (contract no. NMP4-CT-2004-500198), and the
Italian ``Ministero dell'Istruzione, dell Universit\`a e della 
Ricerca'' for financial support
within COFIN 2002. A. Incze also acknowledges support through the EEC
under contract  no. HPRNT-CT-2000-00167 (Univ. Tor Vergata, Roma). 
We would like to thank C. Hogan, M. Palummo, and M. Gatti for useful 
discussions, and N.
Manini for a careful reading of the manuscript. 
Computer facilities at CINECA granted by INFM (Project ID n. 239488704824) are gratefully acknowledged.

\bibliographystyle{apsrev}

%
%

\newpage

 
\begin{table}[htbp]
\begin{tabular}{ccccc} \hline \hline
surface  & ${\alpha}_{1}(^{\circ})$ & $d_{1}$({\AA})  & ${\alpha}_{2}(^{\circ})$ & $d_{2}$({\AA}) \\
\hline
clean    & 19.2                    & 2.33           &     19.2             &   2.33        \\
oxidized & 11      & 3.06    &     20.27     &   2.34   \\
 &    (-43\%)            &  (+31\%)   &      (+6\%)     &    (0.4\%)  \\
\hline \hline
\end{tabular}
\caption{Structural changes (Si-Si buckling angle $\alpha$ and dimer length $d$) at the p(2x2) reconstructed Si(100) surface, subsequent to 0.5 ML oxidation. Subscript $1$ corresponds to the dimer which undergoes oxidation, while subscript $2$ to the dimer which remains clean. In parenthesis, variations in \% with respect to the clean surface. The buckling angles in the case of the clean
surface are the averages of two slightly different angles for the two dimers {\protect\cite{RamstadPRB_51_14504_1995}}.
}
\label{tab:structural}
\end{table}



\end{document}